# An experimental framework for designing document structure for users' decision making: An empirical study of recipes

Rina Kagawa[1], Masaki Matsubara[1], Rei Miyata[2, 4], Takuya Matsuzaki[3], Yukino Baba[1], and Yoko Yamakata[4]

[1] University of Tsukuba, Tsukuba, Japan `kagawa-r@md.tsukuba.ac.jp`
[2] Nagoya University, Nagoya, Japan  [3] Tokyo University of Science, Tokyo, Japan
[4] The University of Tokyo, Tokyo, Japan

Textual documents need to be of good quality to ensure effective asynchronous communication in remote areas, especially during the COVID-19 pandemic. However, defining a preferred document structure (content and arrangement) for improving lay readers' decision-making is challenging. First, the types of useful content for various readers cannot be determined simply by gathering expert knowledge. Second, methodologies to evaluate the document's usefulness from the user's perspective have not been established. This study proposed the experimental framework to identify useful contents of documents by aggregating lay readers' insights. This study used 200 online recipes as research subjects and recruited 1,340 amateur cooks as lay readers. The proposed framework identified six useful contents of recipes. Multi-level modeling then showed that among the six identified contents, *suitable ingredients* or *notes* arranged with a subheading at the end of each cooking step significantly increased recipes' usefulness. Our framework contributes to the communication design via documents.

## 1  INTRODUCTION: FORMALIZING DOCUMENT STRUCTURE FOR LAYPEOPLE'S DECISION-MAKING

Documents allow humans to share diverse and complex information asynchronously to remote areas, especially as the world combats the COVID-19 pandemic. Documents should be organized to achieve their communicative goals [1] (e.g., elementary school textbooks providing young students with age-appropriate knowledge). Any user of a document must be able to retrieve the required information from the document, and documents need to be of good quality so that users can change their thinking and behavior according to the document's communicative goal (Fig. 1(A)). We focused on the quality of the document structure, which consists of content (What should be included in the document?) and arrangement (How should content be organized?) [2].

Readers of documents sometimes make decisions based on their understanding of the documents. This study focused on helping lay readers' decision-making, which is the communicative goal of documents. Documents with principal communicative goals unrelated to readers' decision-making, such as novels [3], are beyond the scope of this study.

When documents aim to help lay readers make decisions, expert knowledge might be insufficient since experts may not understand what lay readers judge to be useful[1]. Furthermore, document structures are often designed by a small number of domain experts [4–6]. Therefore, a framework to defining the structures of these documents for reader decision making is required. Consider recipes as an example of a document that aims to support readers' decision-making. Readers need to make flexible decisions about how to cook based on details provided in recipes (e.g., "You can use tuna instead of salmon.") and their preferences and environments. However, expert cooks may not understand what information amateur cooks find useful.

Developing a method to define a preferred document structure for improving lay readers' decision-making poses two challenges. First, the types of content categories that diverse readers judge useful for decision-making are often empirical and based on their differing understandings and motivations; therefore, these categories might not be determined only by gathering expert knowledge. Second, decisions are not always made immediately after reading[2], making it difficult to directly evaluate documents as useful for decision-making. Moreover, evaluation metrics are also challenging. When the goal of a document is to help readers make decisions, its quality cannot be evaluated solely on the accuracy of the readers' understanding. Furthermore, document evaluations can vary depending on the readers' understandings and motivations, which also makes it difficult to assess the usefulness of documents.

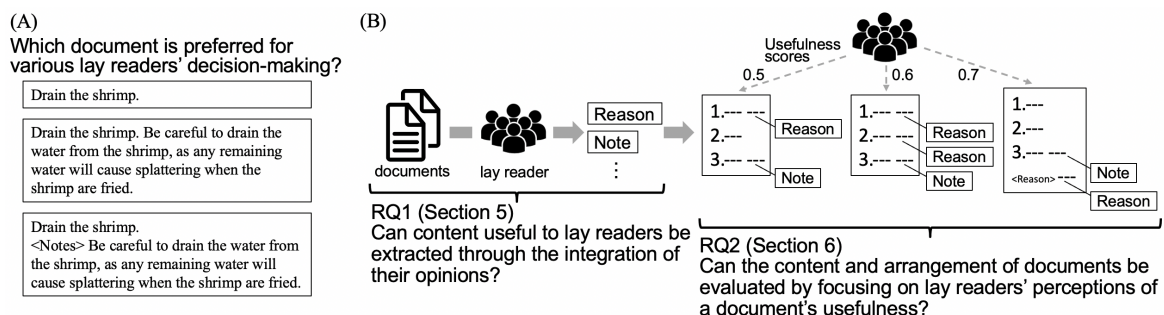

Figure 1. (A) Defining a preferred document structure for various lay readers' decision-making is required. (B) Our proposed framework to define a preferred document structure and research questions.

## 2 OBJECTIVES AND RESEARCH QUESTIONS

The objective of this study is to define a preferred document structure for improving lay readers' decision-making. To achieve this goal, we assume that it is essential to utilize the knowledge of lay readers. As a starting point, we adopted readers' perceptions of a document's usefulness as a metric to evaluate whether that document was useful for their decision-making. The research questions for this study (Fig. 1(B)) are as follows:

- RQ1: Can content useful to lay readers be extracted through the integration of their opinions?
- RQ2: Can the content and arrangement of documents be evaluated by focusing on lay readers' perceptions of a document's usefulness?

---

[1] We regarded understandability, readability, and other metrics and usefulness as interrelated but different concepts. Even if a document is easy to understand, easy to read, and free of contradictions, the document's usefulness for decision-making may not have been properly evaluated unless the reader ultimately found the document useful. Therefore, in this paper, we used usefulness for the reader as a metric (please see Section 2). Understandability, readability, consistency, and so on were not directly discussed.

[2] For example, recipe users sometimes apply content (for example, a cooking process that mentions different ingredients as substitutions) more than a week after having read the recipe.

This study conducted an empirical study. This paper's empirical studies focused on online recipe collections as an example of a document genre whose communicative goal is readers' flexible decision-making. Although recipes are generally well organized procedurally, research has not sufficiently investigated which content (e.g., "why a given process is necessary") and arrangement is useful for diverse readers. Some people might think that recipes should be recorded as images or videos. However, as our preliminary survey, we asked 150 people who have used recipes how important text, images, and videos are in recipes using a 10-point Likert scale (10: important – 1: not important); the average values (standard deviation (SD)) were 9.05 (1.24), 8.63 (1.34), and 6.60 (2.77), respectively. Therefore, the recipe users were deemed to attach more importance to text as information sources than images or videos. Then, this paper's empirical studies focused on text of recipes as an example of textual documents.

This study first proposes an experimental framework for identifying content useful to lay readers of text documents by integrating lay readers' opinions. Then, as an empirical study, we conduct an experiment to actually identify content useful to lay readers using the proposed framework (Section 5). Next, the identified contents' arrangements are evaluated based on the lay readers' perceptions of a document's usefulness, taking into account the readers' diverse understandings and motivations (Section 6).

## 3 RELATED WORKS

### 3.1 Document Formalization

**Research methods.** In recent studies, the frameworks for document formalization were created by aggregating experts' opinions. For example, introduction, method, results, and discussion (IMRaD) [7] has been widely employed for scientific writing. Move analysis has been used to classify the descriptions of research papers [8, 9]. Subjective data, objective data, assessment, and plan (SOAP) [10] is widely accepted worldwide in daily clinical practice as the ideal framework for clinical notes. On the other hand, document formalization based on an aggregation of lay readers' insights has not been discussed. With the growing popularity of crowdsourcing platforms, there are studies of empirical knowledge revealed by lay people's (crowd workers') processing power. For example, lay people showed the characteristics of paintings that differ from painter to painter [11], the characteristics of facial expressions that suggest lying [12], and the shape of stars [13]. Moreover, there are many studies in which lay people modify specialized documents [14] and search for facts based on the documents [15]. Therefore, it is highly possible that lay people would identify useful contents and arrangements from textual documents. We believe that evaluation by crowd workers is particularly appropriate for recipes that are written and read by lay users in their daily lives.

**Research subjects.** Previous studies focused on research papers [7, 8, 9] or clinical notes [10]; these documents are designed to be read by experts. As far as we know, there is few research on document formalization for documents that influence lay readers' flexible decision-making, such as recipes. As the findings relate to the arrangements of documents, headings are effective for proper memory [16–18], but whether the headings are useful for readers' decision-making has not been examined.

**Research purposes.** With the increase in the volume and variety of online documents, not only recipes, information retrieval from web documents [19, 20], search [21, 22], topic and genre estimation [23, 24], and classification [25] of web documents are being actively developed. On the other hand, few studies have examined the preferred document structure based on content and arrangement for online documents [5].

## 3.2 Evaluation Metrics for Texts and Documents

A wide variety of evaluation metrics for the quality of specific documents has been defined. For example, the software requirements specifications must meet 12 requirements [26]. Clinical systematic assessment tool for the understandability and actionability of documents for patient education is used in daily clinical practice [27]. In previous studies, the quality of elementary school textbooks [28], manuals [29–30], documents for technical communication [31], and discussion documents [6] has been discussed. However, the communicative goals of these documents chiefly pertain to the readers' correct understanding or action.

Readability scores [32–36], based on the frequency of words' occurrence, were used to evaluate the texts' readability and applied to code readability for software development [37, 38]. However, these scores are not designed to capture the changes in the document structure, that is, the order of the information presentation. Some studies [39–41] have used psychological modeling of users' sentence processing costs based on their reading time, as measured via eye tracking. However, since these studies could only evaluate the similarities between adjacent words, it is unclear whether they can be applied to more macroscopic evaluations of the effect of the presence or arrangement of the sentences meaning content in the documents on users' decision-making. A psycholinguistic model of sentence processing costs, the concept of "surprisal," was proposed [42], but these costs cannot directly evaluate documents' usefulness. The NASA Task Load Index (NASA-TLX) [43], a measure of mental workload, could be used as an objective measure for document evaluation, but the concept of mental workload focuses on task difficulty or information processing demand, both of which are preprocessing steps in the decision-making process. For this reason, the NASA-TLX is inappropriate for this study.

## 3.3 Research about Recipe Documents

Cooking is a fundamental human activity, deeply associated with people's physical and mental health, and their national and regional cultures and identities [44–46]. Online recipe collections with user-submitted texts have been developed in many countries and languages [47–53]. User-submitted online recipes cannot only contribute to people's physical and mental health in daily life. A large collection of online user-submitted recipes can be a valuable resource for sharing their national and regional cultures and identities for future generations. On the other hand, user-submitted online recipes lack the opportunity for the writer to receive feedback from readers on the quality of the recipe. Therefore, it is worthwhile to experimentally define recipe quality.

In light of rapid technological developments, such as machine learning, some studies have attempted to extract information (e.g., cooking tools [54] or important cooking processes [55]) from recipes. Other studies have generated textual cooking instructions from images [56], the titles and ingredients of recipes [57], or developed recipe-editing systems [58]. As a basis for these studies, an ontology on cooking and ingredients [59–63] and a corpus describing the cooking process in a flow graph [64] are also available. However, these studies focused on the cooking procedures or foods themselves. To our knowledge, no studies have focused on in-depth information beyond ingredients or procedures, such as reasons or substitute of cooking operations. Some studies have pointed out that recipe users may glean in-depth information from recipes, such as arrangements according to taste [65], alternative ingredients [66], and the meanings of cooking procedures [67]. However, whether these descriptions actually improve the usefulness of recipes for readers remains unclear.

## 4 PARTICIPANTS AND MATERIALS

To experimentally realize to define a preferred document structure (content and arrangement) for improving lay readers' decision-making by aggregating lay readers' insights, we had recipes as an example of a document genre whose communicative goal is readers' decision-making. The experiments were performed on a computer, and all participants were recruited from Lancers [68], a Japanese commercial online crowdsourcing platform[3]. All participants connected to the experimental web screen online and participated in the experiment. The study participants responded in a pre-survey that they had cooking experience and reading recipes; all of them did not work in restaurants as chefs but read recipes at least a few times for a month. Therefore, we considered all participants to be amateur cooks as an example of lay readers. In this study, different participants were recruited for each experiment. Any exclusion criteria to recruit participants have been not stetted.

Each experiment utilized several recipes randomly selected from the following recipes: 9,820 online-published recipes written by professional chefs [71] ("professional recipes") and 113,688 online-published recipes submitted by users of recipe collection websites [72] ("amateur recipes"). Details of random selection of recipes are given in the descriptions of each experiment in Section 5 and 6. All recipes were written in Japanese, and all experiments were conducted in Japanese, although they are shown in English in this manuscript for explanatory purposes.

Perl 5.30, Python 3.7.2, NumPy 1.17.2, pandas 0.24.2, NetworkX 2.3, R 4.1.0, brms 2.15.0, lme4 1.1–27, and lattice 0.20–38 were used. This research was approved by the Ethics Committee of Institute of Medicine, University of Tsukuba (permission number: 1522).

## 5 IDENTIFYING CONTENT USEFUL TO LAY READERS

### 5.1 Overview of the framework

To establish a method to define a preferred recipe structure aimed at laypeople's decision-making, this subsection presents an overview of the experimental framework to identify useful contents of recipes. We proposed a three-step experiment to collect, aggregate, and conceptualize the various opinions of lay readers in order to extract categories of useful content for lay readers through the integration of their opinions.

First, to effectively collect the opinions of lay readers, we referred to the cognitive characteristics of people who could recognize objects' characteristics in greater detail when making comparisons [73]. Participants compared two randomly selected recipes. Then, they indicated what made the first recipe more useful than the second, and they also explained what made the second recipe more useful than the first. They used free descriptions in the following form: "It is more useful in terms of the fact that __ for you". We extracted only the unique elements for the descriptions that corresponded to "__" from among those answers. These elements were then called "content candidates."

Next, the content candidates were aggregated according to semantic similarity. This process consists of two stages: defining and capturing similarities between content candidates, and aggregation of the content candidates based on those similarities.

Finally, we assigned the name to each category identified by aggregation; that is, the categories were conceptualized.

### 5.2 Extraction of Categories of Useful Content by Lay Readers

We demonstrated this three-step experiment using recipes and identified the useful categories to lay-readers. In a preliminary study, 40 participants freely described why either the professional or amateur recipes were better. In all, 592 of these descriptions argued that the professional recipes were more useful than the amateur ones, while 558 descriptions claimed that the amateur recipes were more useful. Based on these results, we could not judge whether the professional or amateur recipes were more useful. Therefore, we used a dataset in which the professional and amateur recipes were mixed. The details of each step are shown in Supplement A.

---

[3] Crowdsourcing platforms are online platforms for lay people to accomplish tasks online and is widely used for recruiting participants in academic online experiments since the late 1990s [69, 70].

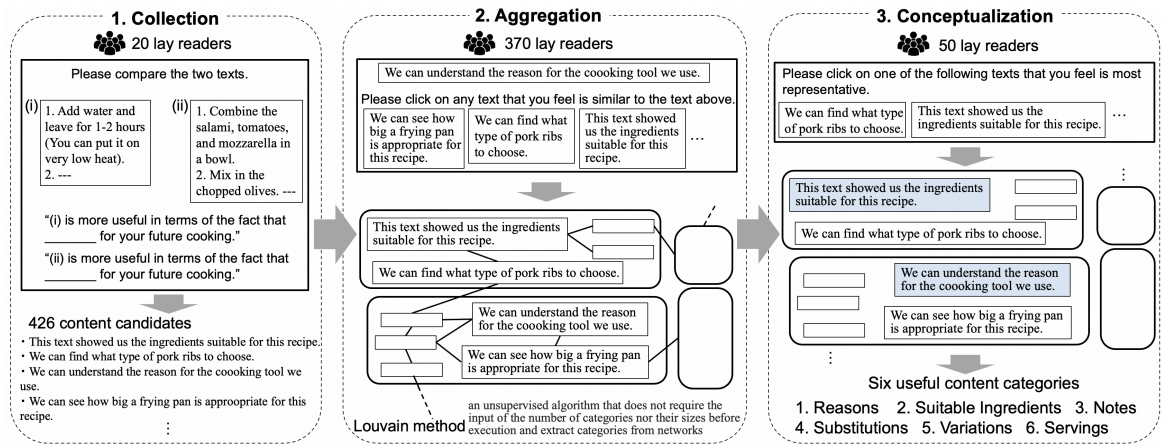

Figure 2. A three-step experimental procedure for extracting categories of useful content.

*5.2.1 Collection*

Twenty participants compared two randomly selected recipes[4]. A total of 134 randomly selected recipes were used. They indicated what made the first recipe more useful than the second, and they also explained what made the second recipe more useful than the first. They used free descriptions in the following form: "It is more useful in terms of the fact that __ for your future cooking" (Fig. 2, left top). Each participant compared two recipes for each task and repeated this task 20 times. The pair of recipes shown varied for each task and participant.

In total, 616 answers were provided. We extracted only the unique elements for the descriptions that corresponded to "__" from among those answers (Fig. 2, left bottom). A total of 426 of these elements were extracted. These elements were then called "content candidates."

*5.2.2 Aggregation*

Next, we aggregated the content candidates according to similarity.

We evaluated the similarities between the content candidates. However, developing technology to automatically judge the similarities between descriptions of varying specificity was difficult. For example, while both of the following two descriptions point out suitable ingredients, the specificity of the words used in the descriptions is greatly different between them.

- "This text showed us the ingredients suitable for this recipe"

---

[4] Each of the dataset of professional recipes and the dataset of amateur recipes has its own three-layer tree structure of sub-categories (i.e., sweets – cakes – chocolate cakes). For both dataset, one recipe originally corresponds to one 3rd-layer sub-category (e.g., chocolate cakes). In this *Collection* step, the two recipes presented each task for each participant were randomly selected from those corresponded to the same 3rd-layer sub-category in the professional and amateur recipe datasets. It is because we assumed that participants would not be able to think deeply about what makes each recipe useful if they compare two very different recipes (e.g., a chocolate cake recipe and a curry recipe). We have to note that we used the professional and amateur recipes were mixed and the two recipes were randomly selected from this mixed dataset; therefore, participants did not necessarily compare the professional recipes with the amateur recipes.

- "We can find what type of pork ribs to choose"

Therefore, the similarities between the content candidates were evaluated manually by asking participants whether the pairs were similar using a majority vote. A total of 370 participants performed 25 tasks each. In each task, we presented the participants with one randomly selected content candidate as a base and another 100 randomly selected content candidates as targets (Fig. 2, middle top). The participants then selected all the target candidates they found to be similar to the base candidate. Pairs of content candidates that four or more study participants judged to be similar were considered similar in this work.

To aggregate the content candidates by similarities, a network was created with each content candidate as a node and a pair of candidates deemed similar as an edge (Fig. 2, middle bottom). We applied the Louvain method, which is a method to extract categories from a network, to this network. The Louvain method is appropriate for aggregating empirical findings because the number of categories obtained by aggregation does not need to be determined in advance. 13 categories were extracted and each category was interpreted as a category required for useful recipes.

*5.2.3 Conceptualization*

Finally, we assigned the name to each category identified by aggregation.

For each category, 50 participants selected the content candidate that best represented the other content candidates within that category by a majority vote (Fig. 2, right top). If we received an equal number of votes for two or more content candidates, one was randomly selected. The content candidate selected for each category was used to name the entire category (Fig. 2, right middle).

The authors divided the 13 categories into two broad types: the recipe's content (e.g., "there is a description of the notes regarding the cooking procedure") and the recipe's expressions (e.g., "cooking instructions are detailed and easy to follow"). For the categories whose names were described as "There is a description of __," "We can understand __," or "__ is written," the "__" was defined as the category about content. The other categories concerned expressions. Of the 13 categories, six were about content (e.g., "the notes regarding the cooking procedure") and seven were about expression. The object of this study was to identify the preferred recipe structure (content and arrangement); therefore, categories of expressions were not discussed hereinafter.

The following six content categories were identified: the reason for the cooking process, information about especially suitable ingredients, the notes regarding the cooking procedure, alternatives to make it delicious, how to arrange (change) the recipe, and how to serve and present dishes. We call these *reasons*, *suitable ingredients*, *notes*, *substitutions*, *variations*, and *servings*, respectively (Fig. 2, right bottom; Table 1).

*5.2.4 Evaluation of the obtained content categories*

Qualitative evaluations were performed to confirm that the lay readers of recipes recognized each content category as a useful item.

As the first qualitative evaluation, we asked 400 participants the following question: "Do you think the recipes that mentioned each content category are useful?" using a 10-point Likert scale (10: Useful – 1: Useless). The mean (standard deviation) of each content category was 8.16 (1.80), 8.48 (1.51), 9.02 (1.22), 8.39 (1.65), 8.33 (1.68), and 6.98 (2.06) for *reasons*, *suitable ingredients*, *notes*, *substitutions*, *variations*, and *servings*, respectively.

For the second qualitative evaluation, we issued a free-form questionnaire that included 200 participants, which sought to ask: "What kind of content is included in the text of useful cooking recipes?" One annotator confirmed whether each content category was included in the answers (multiple categories allowed). For each content category, the percentages of the number of people who mentioned that the content was described in the useful recipes were 17%, 43%, 60%, 26%, 35%, and 5% for *reasons*, *suitable ingredients*, *notes*, *substitutions*, *variations*, and *servings*, respectively. In other words, all content categories were mentioned as useful.

From the above, the content categories obtained by integrating the opinions of lay readers were judged to be appropriate.

Table 1. Examples of sentences for each content category. The sentences annotated with the content are boldfaced and in italics.

| Content category | Sample sentences | The average (SD) of sentences per recipe |
| --- | --- | --- |
| reasons | Turn off the heat and place aluminum foil on the bowl. ***By placing the aluminum foil, you can safely remove the hot bowl from the heat.*** | 0.16 (0.40) |
| suitable ingredients | Fish, 1 slice (80g) ****Spanish mackerel, sea bream, butterfish, etc.*** | 1.16 (1.16) |
| notes | Drain the shrimp. ***Be careful to drain the water from the shrimp, as any remaining water will cause spattering when they are fried.*** | 3.53 (2.61) |
| substitutions | A type of herb characterized by red stems and veins. ***You can also use arugula, marsh, etc.*** | 0.36 (0.75) |
| variations | Add water and leave for 1–2 hours. ***You can put it on very low heat***. | 1.08 (1.43) |
| servings | ***Drain the noodles and place them in the bowl with lettuce and cover liberally with the chicken miso sauce.*** | 0.36 (0.48) |

### 5.3 Discussion

This section demonstrated that content useful to readers could be identified by integrating their opinions. The experiments in Section 5 showed that, depending on the design of the experimental procedure, it is possible to extract the useful contents of documents, which were the empirical insights of the lay readers.

## 6 ANALYSIS OF THE EFFECT OF THE NUMBER AND ARRANGEMENT OF CONTENT CATEGORIES ON THE RECIPES' USEFULNESS

In this section, to establish a method to define a preferred recipe structure aimed at laypeople's decision-making, for each of the useful content categories identified in Section 5, we clarify the arrangement of sentences meaning each content category in cases that the number of the sentences significantly affects recipe usefulness, considering lay-readers' various understandings and motivations.

Section 6.1 details the preprocessing. Section 6.2 details the procedure for collecting data of recipes' usefulness as well as the readers' understandings and motivations. Section 6.3 details the effect of recipe structures on the usefulness of the recipes using the answers acquired in Section 6.2.

Methodologies to evaluate the document's usefulness from the user's perspective have not been established. The purpose of this study was to evaluate the usefulness of the recipes in a reader-focused manner; then, this usefulness was evaluated via questionnaires. The usefulness was calculated by comparing recipes that contained no sentences to indicate the content category and recipe that contained sentences that indicated the content category as per the arrangement.

## 6.1 preprocessing

*6.1.1 Annotation of the content categories*

For the following analysis, we annotated professional recipes, which were not used in the experiments in Section 5, with the content categories because amateur recipes were assumed to include a few sentences indicating the content categories.

From a total of 100 randomly selected professional recipes, we excluded all those that referred to other recipes[5]; this left 86, which we annotated with or without a content category. Two trained annotators tagged the sentences declaiming each content category based on the proposed coding rules[6]. The Kappa coefficient, which is a measure of inter-annotator agreement (1: complete agreement – 0: complete disagreement), was 0.74. In the case of disagreement between the two annotators' decisions, the annotators discussed and uniquely decided the result (Table 1).

*6.1.2 How to make the experimental recipes*

From each annotated recipe ("original recipe"), an experimental recipe ("deleted recipe") was created by deleting a sentence indicating the content category for each content category (Fig. 3 (A)). The deleted recipes contained no sentences indicating the content category. Other experimental recipes ("rearranged recipes," Fig. 3 (B)) were created from each original recipe by reconstructing the sentences meaning the content category into five layout (arrangement) types[7] (A01–A05; Fig. 3 (B)). These five types were determined from a combination of three perspectives: whether they were mentioned at the end of the recipe or at the end of each cooking step, whether line breaks were inserted, and whether headings were used. For each content category, up to five types of rearranged recipes were created based on each original recipe (details are shown in Supplement B). As a result, a total of 36 types of recipe structures (6 content categories × 6 arrangements (original recipe and five types of rearranged recipes)) were discussed in this study.

---

[5] Some recipes published in one online recipe collection refer to other recipes published in the same collection (e.g., a salad recipe refer to another recipe for making the dressing used in that recipe). In this experiment, such recipes were excluded because this experiment focused on completed recipes.
[6] Cooking recipe annotation manual: https://zenodo.org/record/6558280
[7] These arrangements were obtained from a preliminary experiment in which 120 amateur chefs with experience writing recipes were asked to write recipes including each content category without specifying the arrangement.

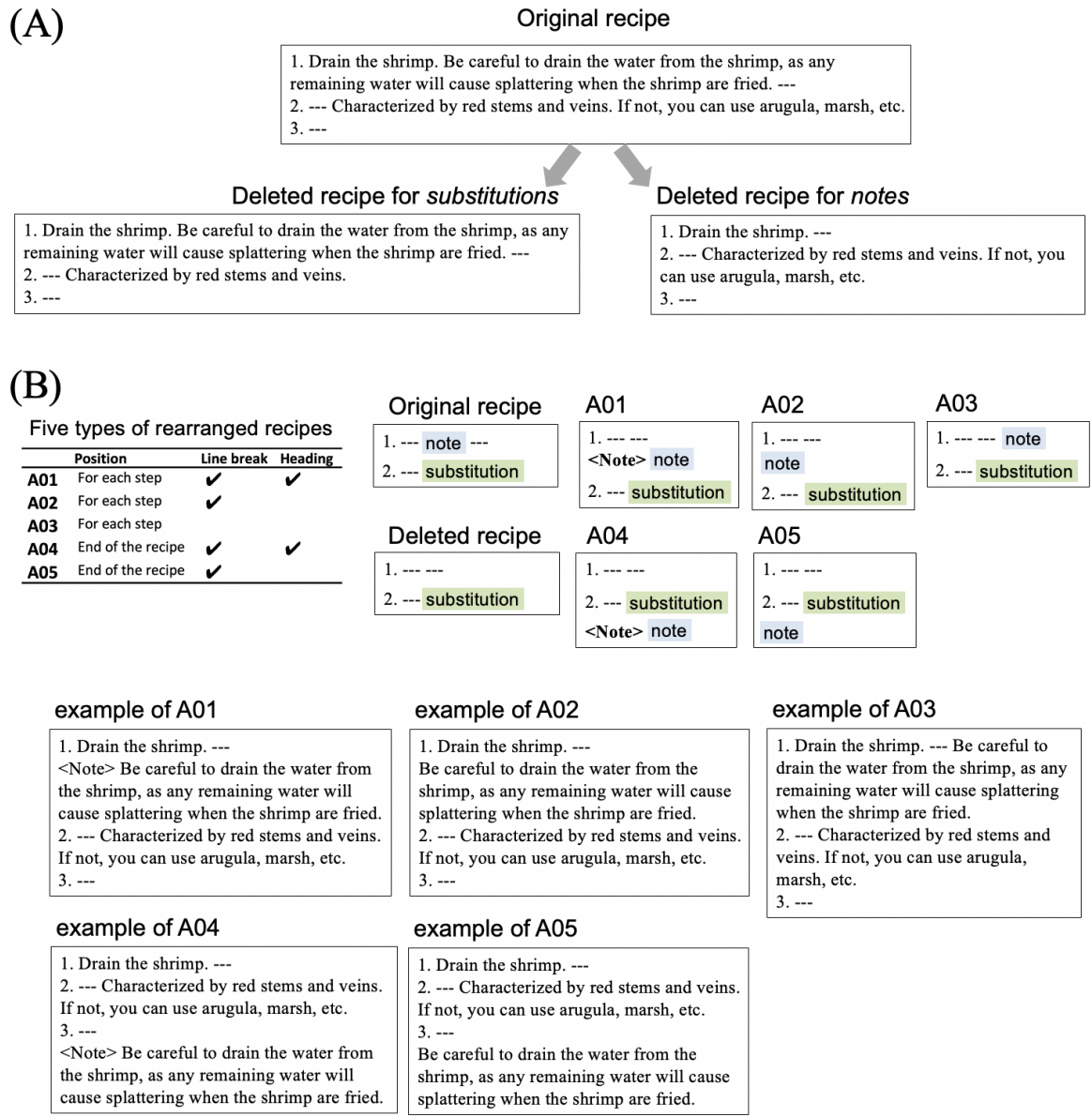

Figure 3. Overview of the experimental recipes. (A) If one original recipe contained multiple content categories, deleted and rearranged recipes for each content category were created. (B) Overview of the original recipes, deleted recipes, and rearranged recipes for *notes* as the content category. Original recipes and rearranged recipes were arrangements discussed in this study, and the deleted recipes were the target of comparison for discussion of arrangements.

## 6.2 Procedures for reader-focused experiments to assess recipe usefulness and readers' background knowledge

In total, 300 participants responded to the survey, which consisted of three steps (Fig. 4).

- **Step 1, Usefulness**: First, participants simultaneously browsed the deleted recipe and one of the corresponding rearranged or original recipes. Using an 11-point Likert scale (10: Useful – 0: Useless)[8], they answered whether each recipe would be useful for future cooking. Each participant repeated this task 25 times, and a pair of a deleted recipe and its corresponding rearranged or original recipe was randomly selected for each task. We estimated that the absolute values of the subjective evaluations of recipe usefulness varied greatly from person to person or recipe to recipe, so we defined the **usefulness score** as the difference between the scores of the original or rearranged recipe and the deleted recipe. That is, if the usefulness score is greater than 0, it means that the usefulness perceived by the reader increased due to the sentence indicating the content category described in that arrangement.
- **Step 2, Understanding**: We sought to examine the readers' understandings as bases for judging the recipe's usefulness; however, there were no metrics for evaluating the readers' understanding of recipes. Therefore, we also developed a questionnaire for Step 2. The participants answered the following three questions for each of the 25 rearranged or original recipes used in Step 1.
    1. **Background knowledge**: How many unknown words are present in the sentences?
    2. **Understandability**: Can you understand the sentences? (10: can – 0: cannot)
    3. **Skill levels**: Can you carry out the instructions that the sentences mean by yourself? (10: can – 0: cannot)
- **Step 3, Information needs**: We examined the kinds of information each reader tended to want while reading recipes, regardless of the recipe. For each content category, we asked each participant once after all other tasks to rate on an 11-point Likert scale how much they wanted to know about that category while they read the recipes.

Combining the usefulness scores, readers' understanding, and information needs, we obtained 7,500 answers. A total of 5,574 answers had no defects in any of the scores and were included in the analysis shown in Section 6.3 (details are shown in Supplement C). The results of the usefulness scores are shown in Table 2.

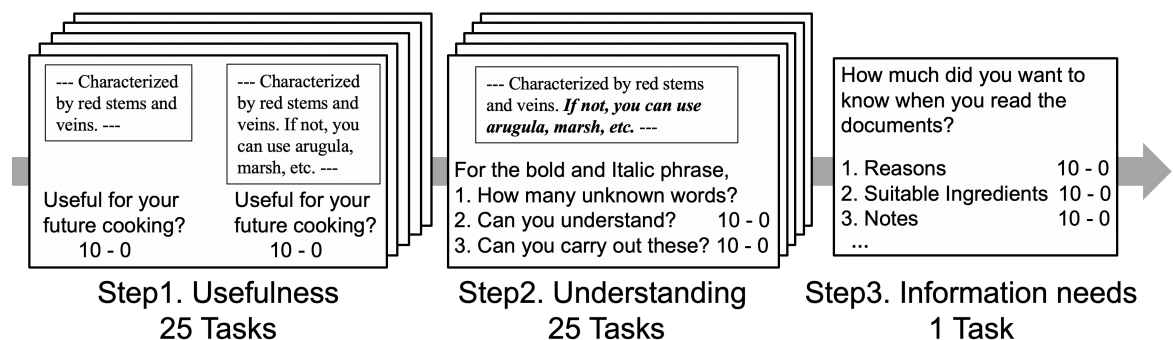

Figure 4. Overview of the experimental procedure used to assess recipe usefulness and readers' background knowledge.

---

[8] The number of Likert scale points to use is still controversial. We referred to research [74] showing that more scale points led to reduced skewness and normal distributions, and we adopted an 11-point scale. The average (SD) of raw data of Likert scale was 7.76 (1.89) and this result showed that the answers did not concentrate on the "middle" choices. Moreover, as described in the manuscript, our analysis did not use the absolute value of the scales, but differences of the values between the deleted recipe and one of the corresponding rearranged or original documents.

Table 2. Averages and standard deviations of usefulness scores for each recipe structure type.

| Content category | Original recipe | Arrangement | | | | |
| --- | --- | --- | --- | --- | --- | --- |
| | | A01 | A02 | A03 | A04 | A05 |
| reasons | 0.41 (0.82) | 0.28 (0.68) | 0.46 (0.78) | 0.37 (1.21) | 0.26 (0.97) | 0.39 (0.94) |
| suitable ingredients | 0.26 (0.69) | 0.46 (0.96) | 0.33 (0.93) | 0.37 (0.80) | 0.34 (0.79) | 0.31 (0.87) |
| notes | 0.66 (1.08) | 0.65 (1.28) | 0.05 (1.85) | 0.68 (1.27) | 0.58 (1.19) | 0.40 (1.31) |
| substitutions | 0.35 (0.90) | 0.31 (0.72) | 0.27 (0.60) | 0.35 (0.90) | 0.37 (0.76) | 0.32 (0.64) |
| variations | 0.32 (0.89) | 0.44 (0.89) | 0.27 (0.81) | 0.45 (1.02) | 0.43 (0.73) | 0.35 (0.85) |
| servings | 0.35 (0.88) | 0.38 (0.70) | 0.40 (0.83) | 0.38 (0.77) | 0.51 (1.06) | 0.44 (1.06) |

### 6.3 Analysis of the effect of the number and arrangement of content categories on recipe usefulness

*6.3.1 Methods*

We investigated the effect of a total of 36 types of recipe structures on recipe usefulness scores using all the answers acquired in the steps described in Section 6.2.

First, we checked which data among participants' understanding and information needs were hierarchical with respect to the usefulness score. The interclass correlation coefficient (ICC) and the design effect (DE)[9] [75] for each data were shown in Table 3. In summary, the usefulness scores were judged to show hierarchy among the presence or absence of unknown words of participants mostly among participants' understandings, information needs because ICCs for all data were under 0.10 and DE was the largest for the presence or absence of unknown words of participants.

Table 3. The interclass correlation coefficient (ICC) and the design effect (DE) for each data.

| | ICC | DE | | ICC | DE |
| --- | --- | --- | --- | --- | --- |
| the presence or absence of unknown words (1: exist – 0: not exist) | 0.03 | 17.3 | the Information needs for *reasons* | 0.01 | 2.17 |
| the number of unknown words | 0.08 | 10.1 | the Information needs for *suitable substitutions* | 0.00 | 1.00 |
| the understandings | 0.00 | 1.00 | the Information needs for *notes* | 0.00 | 1.50 |
| the skill levels | 0.00 | 1.10 | the Information needs for *substitutions* | 0.00 | 1.00 |
| | | | the Information needs for *variations* | 0.08 | 10.0 |
| | | | the Information needs for *servings* | 0.01 | 2.20 |

Therefore, we used multi-level modeling assuming that the usefulness scores showed hierarchy among the presence of participants' unknown words. For each content category, we tried to clarify the arrangement of sentences meaning each content category in cases that the number of the sentences significantly affects recipe usefulness assuming hierarchy among the presence of participants' unknown words. The following six models, Model original and Models A01-A05, were constructed for each content category, and the model with the lowest Widely Applicable Information Criterion (WAIC) value in each content category was used for the analysis (details are shown in Supplement D.1). The model parameters were fitted with four Markov chain Monte Carlo (MCMC) chains with 10,000 iterations and 9,000 burn-in samples and a thinning parameter of one. The six types of models below were compared for each content category.

---

[9] DE is a criterion that takes into account both the average number of data in the group and ICC. $DE = 1 + (k^* - 1) \times ICC$. $k^*$ means the average number of data of the group. An ICC was over 0.1 or a DE of over two suggested that the data were hierarchical.

- Model original: Recipe usefulness is affected by the number of sentences indicating the content category and their arrangements. This effect of the sentences indicating the content category in the original recipe arrangement depends on the presence of unknown words.
- Models A01–A05: Recipe usefulness is affected by the number of sentences indicating the content category and their arrangements. The effect of the sentences indicating the content category rearranged according to the arrangement of A01–A05 depends on the presence of unknown words.

For the six types of models for each content category, we estimated parameters the below equations.

$$usefullness_{ij} = \alpha + \sum_{k=1}^{6} \beta_k n_{ik} + \eta_l^{(z_{ij})} + \gamma_l^{(z_{ij})} n_{il} + e_{ij} \quad (k = 1,..,6)$$

$$e_{ij} \sim N(0, \sigma_e^2)$$

For the usefulness score $usefullness_{ij}$ given to the $i$-th recipe by the $j$-th participants, $\alpha$ was the intercept, $k$ meant the arrangement of the sentences meaning each content category (original, A01, A02, A03, A04, A05), $\beta_k$ ($k = 1,...,6$) was the coefficient for $n_{ik}$, and $n_{ik}$ was the number of descriptions for each of the six content categories. Specifically, we used non-informative prior for $\beta_k$, $\alpha \sim StudentT(3, 0, 2.5)$, and $\sigma_e \sim StudentT(3, 0, 2.5)$ as the prior distributions. $z_{ij} \in \{0, 1\}$ indicated whether there are unknown words in the description of each content category. $\eta_l^{(z_{ij})}$ was the random effect of the arrangement on the intercept for the $l$-th arrangement (e.g., for Model original, original. For Model A01, A01.) of the $i$-th recipe. $\gamma_l^{(z_{ij})}$ was the random effect of the arrangement on the coefficient for $n_{il}$. The models used the same priors as Model base. Non-informative prior was used for the prior distribution of SDs of random effects, and $LKJCholesky(1)$ was the prior of the correlation matrix between $\gamma_l^{(g)}$ and $\eta_l^{(g)}$ for $l \in \{1, ..., 6\}$ and $g \in \{0, 1\}$.

*6.3.2 Results*

The details of the results are shown in Supplement D.2. The results indicated that the number of sentences related to *suitable ingredients* (*mean* = 0.01, 95% credible interval (CI) [0.00, 0.03]) or *notes* (*mean* = 0.03, 95% CI [0.01, 0.04]) significantly increased a recipe's usefulness when the sentences on the contents were arranged with a subheading (such as "<Suitable Ingredients>" or "<Notes>") at the end of each cooking step (arrangement A01).

This study demonstrated that our proposed framework makes it possible to experimentally identify recipe contents and arrangements that significantly increase the usefulness felt by lay readers.

## 6.4 Discussion

This section demonstrated that the recipes in which the number of sentences meaning *suitable ingredients* or *notes* arranged with a subheading at the end of each cooking step significantly increased their usefulness for lay readers were identified experimentally by focusing on lay readers' perceptions of a recipe's usefulness. The result that recipes described in the arrangement using headings were significantly useful was consistent with a previous study [16]. However, this study was the first to experimentally demonstrate this finding using recipes as an example of online documents.

## 7 CONCLUSION AND FUTURE WORKS

To establish a method to define a preferred document structure aimed at laypeople's decision-making, this study demonstrated that content useful to lay readers could be identified by integrating their opinions. Moreover, the contents and arrangements of recipes that significantly increase their usefulness for lay readers were identified experimentally by focusing on lay readers' perceptions of a recipe's usefulness.

The results of this research using recipes showed that if an appropriate experimental procedure (i.e., the three-step framework we proposed) is designed, it is possible to extract useful contents, which are empirical insights of lay readers, and that it is possible to design documents based on these identified useful contents. Our results contribute to practice and research on the design of communication via documents; the obtained preferred document structure can be used not only as a basis for educational applications, such as asking writers to describe documents according to the obtained preferred document structure, but also for research and development of document-authoring tools to generate the preferred document structure.

Our proposed three-step method to identify the useful contents allows both laypeople and experts to be set as users (i.e., it makes it possible to obtain a preferred recipe structure for experts by applying our framework with expert cooks as its users). Our proposed method was also independent of languages. Therefore, our experimental framework has generalizability and applicability. On the other hand, it remains to be seen if our proposed method can be applied to other documents or recipes written in other cultures.

This study adopted the usefulness felt by readers as a metric to evaluate whether a document was useful for readers' decision-making as a first step; therefore, whether the proposed document structure was useful for decision-making in a real-life situation (i.e., real-life cooking situation) was not evaluated. Task-based evaluations in real situations should be incorporated into future research [76]. Background knowledge and other diverse individual differences among readers assumed in this study were obtained only by the developed questionnaire; therefore, developing a more accurate method of estimating readers' background knowledge and other diverse individual differences is future work. Another limitation of this study is that we did not examine how each element was written. Differences in expressions, such as the words used, could affect the usefulness of documents. We plan to conduct a large-scale user experiment by creating documents in which the content is fixed and the expression patterns vary.

## ACKNOWLEDGMENTS

In this paper, we used "Rakuten Dataset" (https://rit.rakuten.com/data_release/) provided by Rakuten Group, Inc. via IDR Dataset Service of National Institute of Informatics. This study was supported by JST-Mirai Program Grant Number JPMJMI19G8 and JSPS KAKENHI Grant Number JP19K19347 and JP21H03993.

**SUPPLEMENT A**

Each experiment was designed so that the work time per participant would be limited to approximately one hour taking into consideration the physical and mental burden on the participants.

**SUPPLEMENT A.1 COLLECTION**

The details of the definitions of the number of participants are as follows:

(20 participants) × (20 tasks (pair of recipes))/(67 pairs of recipes) = 5.97, therefore, approximately five participants, on average, checked each pair of recipes.

**SUPPLEMENT A.2 AGGREGATION**

In this *Aggregation* step, the details of the definitions of the number of participants are as follows:

The number of combinations of 426 content candidates are $_{426}C_2 = 90{,}525$.

370 (participants) × 25 (tasks) × 100 (pairs of content candidates)/90,525 = 10.22, therefore, approximately 10 participants, on average, checked each pair of content candidate.

It is because future research might involve supporting people's efforts to describe the extracted content categories, the threshold value of the similarity (i.e., "Pairs of content candidates that four or more study participants judged to be similar were considered similar") was set so that the number of content categories (see Section 3.1.3) would be between five and nine [ref1], which is the number of elements that humans can retain using short-term memory.

[ref1] Miller, George A. 1956. The magical number seven, plus or minus two: Some limits on our capacity for processing information. Psychol. Rev. 63, 2, 81–97. https://doi.org/10.1037/h0043158

**SUPPLEMENT A.3 CONCEPTUALIZATION**

In this *Conceptualization* step, regarding the categories about content, only "reasons for tools used" was renamed to "reasons for the cooking process," based on this study's intent.

**SUPPLEMENT B**

Table A-1. The number of each type of experimental recipes created for each content category

| Content category | Deleted recipe | Arrangement | | | | |
| --- | --- | --- | --- | --- | --- | --- |
| | | A01 | A02 | A03 | A04 | A05 |
| Reasons | 13 | 13 | 9 | 5 | 13 | 13 |
| Suitable ingredients | 40 | 40 | 40 | 40 | 40 | 40 |
| Notes | 40 | 39 | 40 | 32 | 40 | 40 |
| Substitutions | 21 | 21 | 21 | 21 | 21 | 21 |
| Variations | 44 | 44 | 44 | 17 | 44 | 44 |
| Servings | 30 | 30 | 30 | 1 | 30 | 30 |

When the deleted recipes were created, a minimal expression adjustment was made only when the recipe became grammatically incorrect when the sentences indicating the corresponding content category were deleted. When the rearranged recipes were created, if a rearranged recipe was the same as the original, it was not created. We did not create deleted or rearranged recipes for content categories that were not included in the relevant original recipe.

As a result, the number of each type of experimental recipes created for each content category were shown in the Table A-1. To avoid bias in the number of times each arrangement of content categories was used in the experiments in Section 6.3, we randomly selected 13 experimental documents from all the arrangements of content categories in Table 1 and used them in the experiments in Section 6.3. For the arrangement of content categories Reasons_A02, Reasons_A03, and Servings_A03, for which there were fewer than 13 experimental recipes, all the corresponding experimental recipes were used. Therefore, a total of 444 pairs of the deleted recipe and the original recipe or five types of rearranged recipes were discussed in this analysis.

300 (participants) × 20 (tasks for Step 1 or 2)/444 tasks = 16.89, therefore, approximately over 15 participants, on average, checked each pair of content candidate.

SUPPLEMENT C

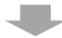

Figure A-1. The details of raw data obtained in Section 6.2. and the data used in Section 6.3.

# SUPPLEMENT D

## SUPPLEMENT D.1  Results for selecting models

The WAIC scores of each model are shown in Table A-2. All $\hat{R}$ were under 1.05.

Table A-2. WAIC values for each model

|  | Reasons | Suitable ingredients | Notes | Substitutions | Variations | Servings |
|---|---|---|---|---|---|---|
| Model original | 14744.9 | 14742.1 | **14674.0** | 14742.0 | **14748.5** | 14752.0 |
| Model A01 | 14746.7 | 14745.4 | 14681.3 | 14742.1 | 14749.0 | 14753.0 |
| Model A02 | **14744.2** | **14741.3** | 14687.2 | 14741.3 | 14750.6 | **14751.3** |
| Model A03 | 14746.5 | 14733.3 | 14680.7 | 14741.7 | 14747.1 | 14751.6 |
| Model A04 | 14745.5 | 14741.7 | 14681.3 | 14746.8 | 14749.7 | 14748.6 |
| Model A05 | 14746.6 | 14742.1 | 14682.3 | **14741.2** | 14750.1 | **14751.3** |

## SUPPLEMENT D.2  Results of parameter fittings

The details of the results for the model with the smallest WAIC for each content category were shown in Table A-3 – A-8.

Table A-3. The estimation results of Model A02 for Reasons

| | Population-Level Effects: | | | | |
|---|---|---|---|---|---|
| | Estimate | Est.Error | l-95% CI | U-95% CI | Rhat |
| Intercept | 0.37 | 0.60 | -1.01 | 1.63 | 1.00 |
| original ($\beta_1$) | 0.00 | 0.01 | -0.01 | 0.02 | 1.00 |
| A01 ($\beta_2$) | -0.01 | 0.01 | -0.03 | 0.00 | 1.00 |
| A02 ($\beta_3$) | -0.03 | 0.78 | -1.83 | 1.67 | 1.00 |
| A03 ($\beta_4$) | 0.01 | 0.01 | -0.01 | 0.04 | 1.00 |
| A04 ($\beta_5$) | -0.01 | 0.01 | -0.03 | 0.00 | 1.00 |
| A05 ($\beta_6$) | -0.01 | 0.01 | -0.02 | 0.01 | 1.00 |
| | Family Specific Parameters: | | | | |
| sigma | 0.91 | 0.01 | 0.89 | 0.92 | 1.00 |
| | Group-Level Effects | | | | |
| | the presence of unknown words in Substitute | | | | |
| sd ($\eta_3$) | 0.63 | 0.99 | 0.01 | 3.43 | 1.01 |
| sd ($\gamma_3$) | 0.81 | 1.04 | 0.01 | 3.78 | 1.00 |
| Corr ($\eta_3$, $\gamma_3$) | 0.02 | 0.63 | -0.96 | 0.98 | 1.00 |

Table A-4. The estimation results of Model A02 for Suitable ingredients

| | Population-Level Effects: | | | | |
|---|---|---|---|---|---|
| | Estimate | Est.Error | l-95% CI | U-95% CI | Rhat |
| Intercept | 0.32 | 0.61 | -1.15 | 1.53 | 1.00 |
| original ($\beta_1$) | -0.02 | 0.01 | -0.03 | -0.00 | 1.00 |
| A01 ($\beta_2$) | **<u>0.01</u>** | 0.01 | **<u>0.00</u>** | **<u>0.03</u>** | 1.00 |
| A02 ($\beta_3$) | -0.01 | 0.01 | -0.02 | 0.01 | 1.00 |
| A03 ($\beta_4$) | 0.05 | 1.12 | -2.36 | 2.55 | 1.00 |
| A04 ($\beta_5$) | -0.00 | 0.01 | -0.02 | 0.01 | 1.00 |
| A05 ($\beta_6$) | -0.00 | 0.01 | -0.02 | 0.01 | 1.00 |
| Family Specific Parameters: | | | | | |
| sigma | 0.91 | 0.01 | 0.89 | 0.92 | 1.00 |
| Group-Level Effects | | | | | |
| the presence of unknown words in Substitute | | | | | |
| sd ($\eta_4$) | 0.71 | 1.02 | 0.01 | 3.45 | 1.00 |
| sd ($\gamma_4$) | 1.12 | 1.28 | 0.08 | 4.73 | 1.00 |
| Corr ($\eta_4$, $\gamma_4$) | -0.08 | 0.60 | -0.98 | 0.95 | 1.00 |

**Table A-5.** The estimation results of Model original for Notes

| Population-Level Effects: | | | | | |
|---|---|---|---|---|---|
| | Estimate | Est.Error | l-95% CI | U-95% CI | Rhat |
| Intercept | 0.31 | 0.63 | -1.27 | 1.46 | 1.00 |
| original ($\beta_1$) | -0.05 | 0.81 | -1.96 | 1.71 | 1.00 |
| A01 ($\beta_2$) | **0.03** | 0.01 | **0.01** | **0.04** | 1.00 |
| A02 ($\beta_3$) | -0.04 | 0.01 | -0.05 | -0.02 | 1.00 |
| A03 ($\beta_4$) | **0.03** | 0.01 | **0.01** | **0.04** | 1.00 |
| A04 ($\beta_5$) | **0.02** | 0.01 | **0.01** | **0.03** | 1.00 |
| A05 ($\beta_6$) | 0.01 | 0.01 | -0.01 | 0.02 | 1.00 |
| Family Specific Parameters: | | | | | |
| sigma | 0.90 | 0.01 | 0.88 | 0.92 | 1.00 |
| Group-Level Effects | | | | | |
| the presence of unknown words in Substitute | | | | | |
| sd ($\eta_1$) | 0.66 | 1.00 | 0.01 | 3.42 | 1.00 |
| sd ($\gamma_1$) | 0.83 | 1.01 | 0.04 | 3.81 | 1.00 |
| Corr ($\eta_1$, $\gamma_1$) | 0.01 | 0.63 | -0.98 | 0.98 | 1.00 |

**Table A-6.** The estimation results of Model A05 for Substitutions

| Population-Level Effects: | | | | | |
|---|---|---|---|---|---|
| | Estimate | Est.Error | l-95% CI | U-95% CI | Rhat |
| Intercept | 0.32 | 0.71 | -1.24 | 1.79 | 1.00 |
| original ($\beta_1$) | -0.01 | 0.01 | -0.02 | 0.00 | 1.00 |
| A01 ($\beta_2$) | -0.01 | 0.01 | -0.02 | 0.01 | 1.00 |
| A02 ($\beta_3$) | -0.01 | 0.01 | -0.03 | 0.00 | 1.00 |
| A03 ($\beta_4$) | -0.00 | 0.01 | -0.02 | 0.01 | 1.00 |
| A04 ($\beta_5$) | -0.00 | 0.01 | -0.02 | 0.01 | 1.00 |

| A05 ($\beta_6$) | -0.05 | 0.72 | -2.08 | 1.50 | 1.00 |
| --- | --- | --- | --- | --- | --- |
| Family Specific Parameters: | | | | | |
| sigma | 0.91 | 0.01 | 0.89 | 0.92 | 1.00 |
| Group-Level Effects | | | | | |
| the presence of unknown words in Substitute | | | | | |
| sd ($\eta_6$) | 0.70 | 1.07 | 0.01 | 3.67 | 1.01 |
| sd ($\gamma_6$) | 0.75 | 1.14 | 0.01 | 4.02 | 1.00 |
| Corr ($\eta_6$, $\gamma_6$) | -0.04 | 0.63 | -0.98 | 0.97 | 1.00 |

**Table A-7.** The estimation results of Model original for Variations

| Population-Level Effects: | | | | | |
| --- | --- | --- | --- | --- | --- |
| | Estimate | Est.Error | l-95% CI | U-95% CI | Rhat |
| Intercept | 0.37 | 0.60 | -1.03 | 1.65 | 1.00 |
| original ($\beta_1$) | -0.03 | 0.71 | -1.57 | 1.54 | 1.00 |
| A01 ($\beta_2$) | 0.00 | 0.01 | -0.01 | 0.02 | 1.00 |
| A02 ($\beta_3$) | -0.01 | 0.01 | -0.03 | 0.00 | 1.00 |
| A03 ($\beta_4$) | -0.00 | 0.01 | -0.02 | 0.01 | 1.00 |
| A04 ($\beta_5$) | 0.00 | 0.01 | -0.01 | 0.02 | 1.00 |
| A05 ($\beta_6$) | -0.00 | 0.01 | -0.01 | 0.01 | 1.00 |
| Family Specific Parameters: | | | | | |
| sigma | 0.91 | 0.01 | 0.89 | 0.92 | 1.00 |
| Group-Level Effects | | | | | |
| the presence of unknown words in Substitute | | | | | |
| sd ($\eta_1$) | 0.62 | 0.94 | 0.01 | 3.31 | 1.00 |
| sd ($\gamma_1$) | 0.64 | 0.90 | 0.01 | 3.23 | 1.00 |
| Corr ($\eta_1$, $\gamma_1$) | 0.00 | 0.63 | -0.98 | 0.98 | 1.00 |

**Table A-8_1.** The estimation results of Model A02 for Servings

| Population-Level Effects: | | | | | |
| --- | --- | --- | --- | --- | --- |
| | Estimate | Est.Error | l-95% CI | U-95% CI | Rhat |
| Intercept | 0.38 | 0.53 | -0.82 | 1.58 | 1.00 |
| original ($\beta_1$) | -0.00 | 0.01 | -0.01 | 0.01 | 1.00 |
| A01 ($\beta_2$) | 0.00 | 0.01 | -0.01 | 0.02 | 1.00 |
| A02 ($\beta_3$) | -0.01 | 0.66 | -1.64 | 1.33 | 1.00 |
| A03 ($\beta_4$) | 0.00 | 0.03 | -0.05 | 0.07 | 1.00 |
| A04 ($\beta_5$) | 0.01 | 0.01 | -0.00 | 0.02 | 1.00 |
| A05 ($\beta_6$) | 0.00 | 0.01 | -0.01 | 0.02 | 1.00 |
| Family Specific Parameters: | | | | | |
| sigma | 0.91 | 0.01 | 0.89 | 0.92 | 1.00 |
| Group-Level Effects | | | | | |

| the presence of unknown words in Substitute | | | | | |
|---|---|---|---|---|---|
| sd ($\eta_3$) | 0.60 | 0.92 | 0.01 | 3.22 | 1.00 |
| sd ($\gamma_3$) | 0.64 | 0.97 | 0.01 | 3.42 | 1.00 |
| Corr ($\eta_3$, $\gamma_3$) | -0.04 | 0.62 | -0.98 | 0.97 | 1.00 |

**Table A-8_2.** The estimation results of Model A05 for Servings

| Population-Level Effects: | | | | | |
|---|---|---|---|---|---|
| | Estimate | Est.Error | l-95% CI | U-95% CI | Rhat |
| Intercept | 0.41 | 0.61 | -0.81 | 1.72 | 1.00 |
| original ($\beta_1$) | -0.00 | 0.01 | -0.01 | 0.01 | 1.00 |
| A01 ($\beta_2$) | 0.00 | 0.01 | -0.01 | 0.02 | 1.00 |
| A02 ($\beta_3$) | 0.00 | 0.01 | -0.01 | 0.02 | 1.00 |
| A03 ($\beta_4$) | 0.01 | 0.03 | -0.04 | 0.07 | 1.00 |
| A04 ($\beta_5$) | 0.01 | 0.01 | -0.00 | 0.03 | 1.00 |
| A05 ($\beta_6$) | 0.01 | 0.63 | -1.29 | 1.57 | 1.00 |
| Family Specific Parameters: | | | | | |
| sigma | 0.91 | 0.01 | 0.89 | 0.92 | 1.00 |
| Group-Level Effects | | | | | |
| the presence of unknown words in Substitute | | | | | |
| sd ($\eta_6$) | 0.61 | 0.88 | 0.01 | 3.18 | 1.00 |
| sd ($\gamma_6$) | 0.66 | 0.92 | 0.01 | 3.35 | 1.01 |
| Corr ($\eta_6$, $\gamma_6$) | 0.02 | 0.63 | -0.97 | 0.98 | 1.00 |